\documentclass[%
 reprint,
 amsmath,amssymb,
 aps,
]{revtex4-1}

\usepackage{graphicx}
\usepackage{dcolumn}
\usepackage{bm}
\usepackage{float} 

\newcommand*{\affaddr}[1]{#1} 
\newcommand*{\affmark}[1][*]{\textsuperscript{#1}}

\begin{document}

\preprint{APS/123-QED}

\title{Searching for New Heavy Neutral Gauge Bosons using Vector Boson Fusion Processes at the LHC} 

\author{
Andr\'es Fl\'orez\affmark[2],  Alfredo Gurrola\affmark[1], Will Johns\affmark[1], Young Do Oh\affmark[3], Paul Sheldon\affmark[1], Dylan Teague\affmark[1], and Thomas Weiler\affmark[1]\\
\affaddr{\affmark[1] Department of Physics and Astronomy, Vanderbilt University, Nashville, TN, 37235, USA}\\
\affaddr{\affmark[2] Physics Department, Universidad de los Andes, Bogot\'a, Colombia}\\
\affaddr{\affmark[3] Department of Physics, Kyungpook National University, Deagu, Korea}\\
}

\date{\today}

\begin{abstract}

New massive resonances are predicted in many extensions to the Standard Model (SM) of particle physics and constitutes one of the most promising searches for new physics at the LHC. We present a feasibility study to search for new heavy neutral gauge bosons using vector boson fusion (VBF) processes, which become especially important as the LHC probes higher collision energies. In particular, we consider the possibility that the discovery of a $Z'$ boson may have eluded searches at the LHC. The coupling of the $Z'$ boson to the SM quarks can be small, and thus the $Z'$ would not be discoverable by the searches conducted thus far. In the context of a simplified phenomenological approach, we consider the $Z'\to\tau\tau$ and $Z'\to\mu\mu$ decay modes to show that the requirement of a dilepton pair combined with two high $p_{T}$ forward jets with large separation in pseudorapidity and with large dijet mass is effective in reducing SM backgrounds. The expected exclusion bounds (at 95\% confidence level) are $m(Z') < 1.8$ TeV and $m(Z') < 2.5$ TeV in the $\tau\tau j_{f}j_{f}$ and $\mu\mu j_{f}j_{f}$ channels, respectively, assuming 1000 fb$^{-1}$ of 13 TeV data from the LHC. The use of the VBF topology to search for massive neutral gauge bosons provides a discovery reach with expected significances greater than 5$\sigma$ (3$\sigma$) for $Z'$ masses up to 1.4 (1.6) TeV and 2.0 (2.2) TeV in the $\tau\tau j_{f}j_{f}$ and $\mu\mu j_{f}j_{f}$ channels.
\end{abstract}

\pacs{Valid PACS appear here}
\maketitle


\section{\label{sec:level1}Introduction}

The discovery of a Higgs boson with mass of about 125 GeV in 2012~\cite{Aad20121,Chatrchyan201230} at the CERN Large Hadron Collider (LHC) has initiated a new era of excitement in particle physics aimed at the understanding of the full nature of electroweak symmetry breaking. In particular, the question of whether this new boson is solely responsible for electroweak symmetry breaking and the origin of mass is a central thrust of the physics program at the LHC. 

The Standard Model (SM) of particle physics has been successful at explaining a wide range of experimental phenomena. However, there are many open questions it fails to answer. Similar to the introduction of the Higgs mechanism to explain the SU(2)$\times$U(1) symmetry breaking of the SM, several extensions to the SM model have been proposed to address its incompleteness. 
For example, models of extra dimensions, such as the Randall-Sundrum (RS) model~\cite{Randall:1999ee}, were introduced to solve the hierarchy problem by postulating that gravity is weaker compared to the other fundamental forces because it can propagate in extra spatial dimensions. While the extent of these new physics models is vast and their physics motivation often different,  a common theme is the manifestation of new TeV scale neutral gauge bosons ($Z'$) that can be probed at the LHC.

At the LHC, the ATLAS~\cite{Aad:2008zzm} and CMS~\cite{Chatrchyan:2008aa} experiments have an extensive physics program to search for $Z'$. A widely used model in these searches is the Sequential Standard Model (SeqSM), which predicts a spin-1 neutral boson ($Z'_{SeqSM}$) with SM-like couplings~\cite{SeqSMZprime}. Results of direct searches for high mass dilepton resonances in proton-proton collisions at $\sqrt{s}=7$, 8, and 13 TeV exclude $Z'_{SeqSM}$ masses below 3.75 and 2.1 TeV in the $\mu\mu$ and $\tau\tau$ channels, respectively~\cite{ATLASZprimeMuMuPAS,CMSZprimeMuMuPAS,ATLASZprimeTauTau,CMSZprimeTauTauPAS}. The strategy pursued in those analyses is a simple and robust dilepton selection, targeting production via Drell-Yan (DY) processes of order $\alpha_{EW}^{1}$ (i.e. $q\bar{q} \to Z'+0j$),  with high signal acceptance that produces a "bump" (narrow in $\mu\mu$ and broad in $\tau\tau$) in the reconstructed invariant mass spectrum of lepton pairs that sits above a smooth and steeply falling background distribution. 

\begin{figure}[H]
 \begin{center} 
 \includegraphics[width=0.5\textwidth, height=0.35\textheight]{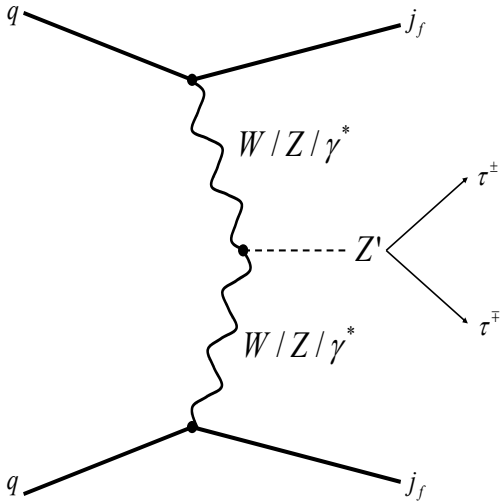}
 \end{center}
 \caption{Feynman diagram depicting pure electroweak VBF production of a $Z'$ particle and two forward jets.}
 \label{fig:feyn}
 \end{figure}

The focus of this paper is to propose new searches for $Z'$ bosons at the LHC using events produced through vector boson fusion (VBF) processes (Figure~\ref{fig:feyn}). The tagging of events produced though VBF processes has been proposed by some of the present authors as an effective experimental tool for dark matter (DM) and electroweak supersymmetry (SUSY) searches at the LHC~\cite{VBF1, DMmodels2,VBFSlepton,VBFStop}. In particular, it has been shown in ~\cite{VBF1, DMmodels2,VBFSlepton,VBFStop} that VBF processes become increasingly more important for probing large mass scales due to collinear logarithmic enhancements in the production cross-sections. This characteristic is particularly interesting in light of the recent observation of a $> 3\sigma$ excess (local significance) near $m_{\gamma\gamma} = 750$ GeV in the high-mass diphoton searches at ATLAS and CMS. Although the significance of this bump has decreased with the 13 TeV data acquired this year~\cite{Khachatryan:2016yec, Aaboud:2016tru}, it has motivated exploring the effectiveness of the VBF topology for TeV scale $Z'$ searches. Additionally, the VBF topology provides significant reduction of the QCD multijet background in SUSY and DM searches~\cite{VBF2,CMSVBFDM}, which is an important characteristic that can also be utilized in $Z'$ searches characterized by large QCD multijet backgrounds, such as $Z'\to\tau\tau$.

Although there are many ways that a new heavy neu-

\noindent tral gauge boson may arise from extensions to the SM, in order to allow for a broad and generic discussion about the importance of VBF $Z'$ under various model assumptions, we consider a ``simplified phenomenological approach" where the $Z'$ mass and the $Z'$ couplings to the SM quarks ($g_{Z'qq}$) and vector bosons ($g_{Z'VV}$) are free parameters~\cite{HeavyVectorTriplets}. To highlight the usefulness of the VBF topology in $Z'$ searches with large QCD backgrounds as well as those with clean signatures, we consider $Z'$ decays to $\tau\tau$ and $\mu\mu$. In the VBF $Z'\to\tau\tau$ study, we focus on the final state where both $\tau$ leptons decay to hadrons ($\tau_{h}$) since it provides the largest $\tau\tau$ branching fraction (42\%) and is expected to produce better sensitivity compared to final states with semi-leptonic decays of $\tau$ leptons.

\section{Samples and simulation}

The dominant sources of background in these studies are production of top quark pairs ($t\bar{t}$) and $Z$/$W$ bosons with associated jets (referred to as $Z$+jets, $W$+jets, and more generally as $V$+jets). The $Z$+jets background is characterized by two real prompt leptons from the $Z$ boson decay in addition to two jets, either from initial state radiation (processes of order $\alpha_{EW}^{1}\alpha_{QCD}^{2}$) or from pure electroweak processes. The W+jets background is only important in the $\tau\tau$ study and satisfies the selection criteria when a jet is misidentified as a $\tau_{h}$ and two additional jets (e.g. initial state radiation) satisfy the VBF criteria. Background from $t\bar{t}$ events is usually accompanied by one or two b quark jets, in addition to real prompt leptons. 
Signal and background samples were generated with MadGraph (v2.2.3) \cite{MADGRAPH}, considering proton-proton

 \begin{figure}[H]
 \begin{center} 
 \includegraphics[width=0.45\textwidth, height=0.35\textheight]{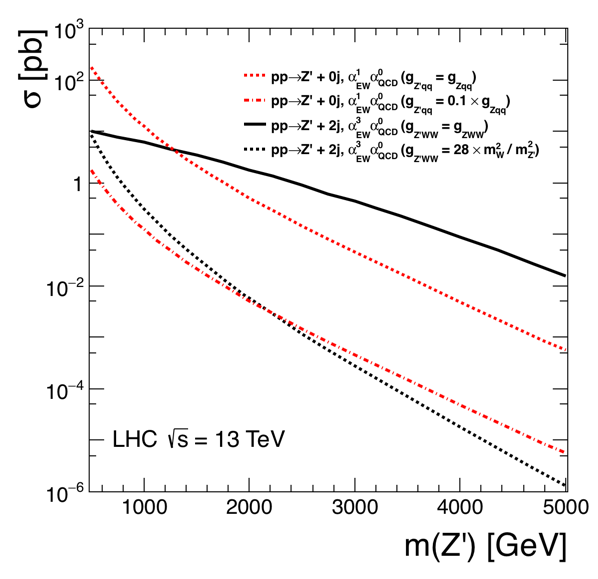}
 \end{center}
 \caption{The $Z'$ production cross-section as a function of mass.}
 \label{fig:XSections}
 \end{figure} 

\noindent  beams with $\sqrt{s}=13$ TeV. PYTHIA (v6.416) \cite{Sjostrand:2006za} was used for the hadronization process, while Delphes  (v3.3.2) \cite{deFavereau:2013fsa} was used to simulate detector effects (CMS configuration was used). The $Z$/$W$($\to \ell\ell$/$\ell\nu$)+jets background events were generated with up to four associated jets, inclusive 
in $\alpha_{EW}$ and $\alpha_{QCD}$ (e.g. $Z+2j$ includes both pure electroweak diagrams as well as diagrams with two jets from initial state radiation). The MLM algorithm \cite{MLM} was used for jet matching and jet merging, which optimizes two variables (xqcut and qcut) related to the jet definition. The xqcut variable defines the minimal distance between partons at MadGraph level. The qcut variable defines the minimum energy spread for a clustered jet in PYTHIA. In order to determine appropriate xqcut and qcut values, the distribution of the differential jet rate was required to smoothly transition between events with $N$ and $N+1$ jets. The jet matching and merging studies resulted in an optimized xqcut of 20 and qcut of 40. At the MadGraph level, leptons were required to have a $p_{T} (\ell) > 10$ GeV and $|\eta (\ell)| < 2.5$, while jets were required to have a minimum $p_{T}> 20$ GeV and $|\eta| < 5.0$. 

The signal samples were produced considering only pure electroweak production of $Z'$ and two associated jets: $pp\to Z'jj$ with $\alpha_{QCD}^{0}$. As mentioned above, we considered a ``simplified phenomenological approach" and allowed decays to either $\tau\tau$ or $\mu\mu$. We scanned $m(Z')$ ranging from 1 TeV to 5 TeV, in steps of 250 GeV. The $Z'$ coupling to quarks is defined as $g_{Z'qq} = \kappa_{q} \times g_{Zqq}$, where $g_{Zqq}$ is the SM $Z$ boson coupling to quarks and $\kappa_{q}$ is a ``modifier" for the coupling. For example, if $\kappa_{q} = 1$ we recover the SeqSM scenario. The $Z'$ coupling to the SM vector bosons can be similarly defined as $g_{Z'VV} = \kappa_{V} \times g_{ZVV}$, where $g_{ZVV}$ is the SM $Z$ boson coupling to the SM vector bosons and $\kappa_{V}$ is a ``modifier" for the coupling. Figure~\ref{fig:XSections} shows the $Z'$ production cross-section as a function of mass for both "standard" DY-like production, which are consistent with the values used by ATLAS and CMS in~\cite{ATLASZprimeMuMuPAS,CMSZprimeMuMuPAS,ATLASZprimeTauTau,CMSZprimeTauTauPAS}, and pure electroweak VBF production. The VBF $Z'$ production cross-section surpasses that of the DY process at approximately $m(Z')=1.25$ TeV for $\kappa_{V} = 1$.

\begin{table}[H]
\begin{center}
\caption {Event selection criteria used for the VBF $Z'(\to\tau\tau$/$\mu\mu)$ + $j_{f}j_{f}$ channels.}
\label{tab:selections}
\begin{tabular}{ l  c c}\hline\hline
Criterion & $\tau_{h}\tau_{h} j_{f}j_{f}$ &  $\mu\mu j_{f}j_{f}$\\
 \hline
  \multicolumn{3}{ c }{{\bf Central Selections}} \\
   \hline
  $|\eta(\tau_{h}/\mu)|$ &  $< 2.5$ &  $< 2.5$\\  
  $p_{T}(\tau_{h}/\mu)$ & $> 60$ GeV & $>  30$ GeV\\
  $N(\tau_{h}/\mu)$ & $\ge 2$ & $\ge 2$ \\
  $Q(\tau_{h,1}/\mu_{1})\cdot Q(\tau_{h,2}/\mu_{2})$ & $< 0$ & $< 0$\\
  $\Delta R(\tau_{h1}/\mu_{1},\tau_{h2}/\mu_{2})$ & $> 0.3$ & $> 0.3$ \\ 
   N(b-jets)  & 0 & 0\\
   $p_{T}^{miss}$ & $> 30$ GeV & -- \\
   \hline
   \multicolumn{3}{ c }{{\bf VBF Selections}} \\
   \hline
  $p^{lead}_{T}(jet)$ & 50 GeV & 30 GeV \\
   $|\eta^{lead} (jet)|$ & $< 5.0$  & $< 5.0$\\
   $p^{sub-lead}_{T}(jet)$ & 50 GeV & 30 GeV \\
   $|\eta^{sub-lead} (jet)|$ & $< 5.0$  & $< 5.0$\\
   $\Delta R(\tau_{h}/\mu, jet)$ & $> 0.4$ & $> 0.4$ \\
   $\eta(j_{1})\cdot \eta(j_{2})$ & $< 0$ & $< 0$ \\
   $|\Delta \eta (j_{1}, j_{2})| $ & $> 4.0$ & $> 4.0$\\
   $m_{jj}$ & $> 1.0$ TeV & $> 0.5$ TeV \\
   \hline\hline
 \end{tabular}
\end{center}
\end{table}

\noindent If the $Z'$ couples to the SM vector bosons, the $Z'$ decay width is given by $\frac{g_{Z'VV}^{2}cos^{2}\theta_{\omega}m_{Z'}^{5}}{192\pi m_{W}^{4}}$, where $\theta_{\omega}$ is the weinberg angle and $m_{W}$ is the mass of the SM $W$ boson. Therefore, the ``region of validity" (i.e. the decay width must be smaller than the $Z'$ mass) is defined by the maximal coupling $g_{Z'VV}^{max} = (5.3 \times m_{W} / m_{Z'})^{2}$. For this reason, in the remainder of this paper the coupling $g_{Z'VV}$ is defined as $g_{Z'VV} = \kappa_{V} \times g_{Z'VV}^{max}$, where $\kappa_{V} \le 1$. As shown in Figure 2, while the VBF $Z'$ cross-section cannot exceed the DY $Z'$ cross-section for $g_{Z'VV} = g_{Z'VV}^{max}$ and $\kappa_{q} = 1$, the VBF $Z'$ search can become the most important mode for discovery when $\kappa_{q}$ is small (how small will be shown in subsequent sections). This is particularly interesting since there has been no evidence for a TeV scale $Z'$ at the LHC. The $Z'$ boson may have eluded the CMS and ATLAS experiments thus far because $g_{Z'qq}$ is small, and thus would not be discoverable by the ``standard" DY $Z'$ searches conducted thus far. On the other hand, it is important to point out that even if the DY $Z'$ cross-section dominates when $\kappa_{q} \approx 1$, a VBF $Z'$ search remains a key part of the $Z'$ search program at the LHC in order to establish the coupling of the $Z'$ boson to the vector bosons of the SM, which is important to assess the correct physics model should there be potential evidence for discovery (the recent 750 GeV diphoton bump is a good example). Finally, we also point out that our simplified phenomenological approach covers scenarios/models where the $Z'$ indirectly couples to the SM vector bosons through $Z'$-$Z$ mixing. The mixing can absorbed in the definition of $g_{Z'VV}^{max}$ and/or $\kappa_{V}$. For completeness, it is noted the cross-sections for VBF Higgs production in~\cite{Aad20121,Chatrchyan201230} have also been reproduced as a cross-check of the MadGraph calculations for VBF $Z'$.

\section{Event selection criteria}

The VBF topology is characterized by two high $p_{T}$ forward jets ($j_{f}$), with large pseudorapidity gap, located 
in opposite hemispheres of the detector, and TeV scale dijet invariant masses. Similar to the VBF SUSY and VBF DM searches in~\cite{VBF2,CMSVBFDM}, an advantage of this topology is the reduction of the QCD multijet background particularly important in searches with $\tau$ leptons, which have relatively high jet$\to\tau_{h}$ misidentification rates. However, 
a VBF $Z'$ search differs fundamentally from those analyses in the following ways: $(i)$ since SUSY particles must be produced in pairs under the assumption of R-parity conservation, the VBF processes considered in~\cite{VBF1, DMmodels2,VBFSlepton,VBFStop,VBF2,CMSVBFDM}  
occur through t-channel diagrams, while VBF $Z'$ proceeds via s-channel (Figure~\ref{fig:feyn}), resulting in lower average $p_{T}$ for the jets; $(ii)$ VBF $Z' \to \ell\ell$ contains leptons with 
higher $p_{T}$ (TeV scale). These features mean different search philosophies, including e.g. the $\tau_{h}$ identification strategy and the choice of the experimental trigger. For example, while the VBF SUSY searches profit from the use of a VBF dijet + $p^{miss}_{T}$ trigger ($p^{miss}_{T}$ is the missing transverse momentum in an event) to select low $p_{T}$ leptons without bias, the high $p_{T}$ leptons from $Z'$ decays allow the use of single lepton or dilepton triggers to maintain unbiased VBF dijet and $p^{miss}_{T}$ cuts targeting the lower jet $p_{T}$ and $p^{miss}_{T}$ values typical of s-channel VBF $Z'$ production. While discovery of either type of new physics would be an incredible achievement, the difference between t-channel and s-channel provides a clean

 \begin{figure}[H]
 \begin{center} 
 \includegraphics[width=0.5\textwidth, height=0.35\textheight]{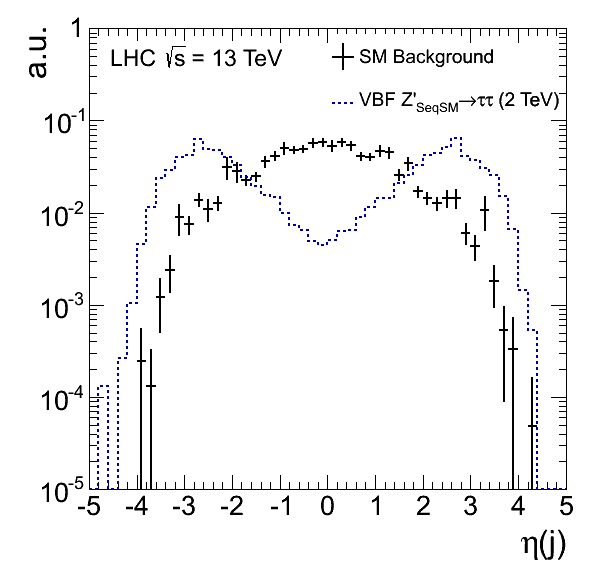}
 \end{center}
 \caption{$\eta$ distribution of the highest $p_{T}$ jet for the sum of SM backgrounds and the VBF $Z'\to\tau\tau$ signal benchmark point with $m(Z')=2$ TeV.}
 \label{fig:jeteta}
 \end{figure}

\noindent approach to reveal the difference between the two models.

\begin{figure}[H]
 \begin{center} 
 \includegraphics[width=0.5\textwidth, height=0.35\textheight]{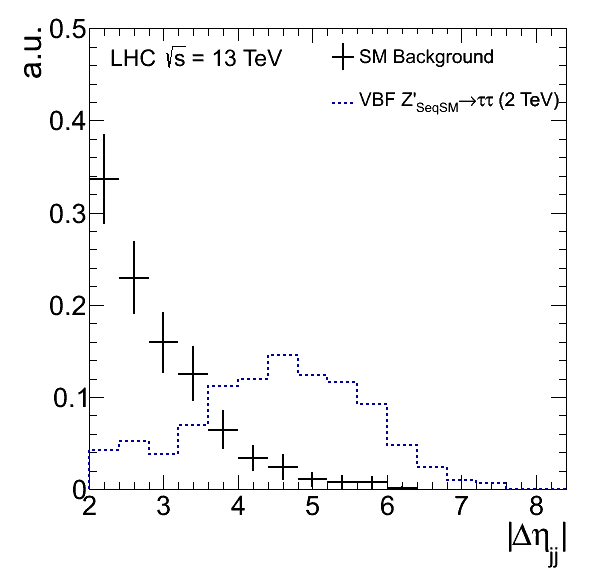}
 \end{center}
 \caption{$|\Delta\eta|$ between the two leadings jets for the sum of SM backgrounds and the VBF $Z'\to\tau\tau$ signal benchmark point with $m(Z')=2$ TeV.}
 \label{fig:deta}
 \end{figure}

The event selection criteria used in both the $\tau_{h}\tau_{h} j_{f}j_{f}$ and $\mu\mu j_{f}j_{f}$ channels is summarized in Table~\ref{tab:selections}. In 
the $\tau_{h}\tau_{h} j_{f}j_{f}$ channel, we select events with two oppositely charged $\tau_{h}$ candidates ($Q(\tau_{h1})\cdot Q(\tau_{h2}) < 0$), each with $p_{T}(\tau_{h}) >60$ GeV and $|\eta(\tau_{h})| < $ 2.5, that must be well separated in $\eta-\phi$ space by requiring $\Delta R = \sqrt{\Delta \eta^{2}(\tau_{h,1}, \tau_{h,2}) + \Delta \phi^{2}(\tau_{h,1}, \tau_{h,2})} > 0.3$. The requirement on  $\eta(\tau_{h})$ bounds the selected $\tau_{h}$ candidates to be 
within the tracker acceptance region of the detector. Contamination from $t\bar{t}$ events is largely suppressed by vetoing events containing jets, with $p_{T} > 20$ GeV, identified as b quarks. To further reduce possible contamination from QCD multijet and $Z \to \textrm{ee}$ +jets backgrounds, events must have $p_{T}^{miss} > 30$ GeV. 
In Table~\ref{tab:selections} we refer to the aforementioned criteria as central selections.  
In order to impose a stringent VBF selection aimed at the reduction of QCD multijet and background processes with two jets from initial state radiation (e.g. $pp\to Z jj$ order $\alpha_{EW}^{1}\alpha_{QCD}^{2}$), a minimum $p_{T}$ threshold of 50 GeV is used for the leading and sub-leading jets. The selected jets and $\tau_{h}$ candidates must be well separated by requiring 
$\Delta R(j, \tau_{h}) > 0.4$. The $\eta$ distribution of the highest $p_{T}$ jet for the signal and sum of SM backgrounds is shown in Figure~\ref{fig:jeteta}. Figure~\ref{fig:deta} displays the difference in pseudorapidity, $|\Delta\eta_{jj}|$, between the two leading jets. 
The requirement that the two jets be in opposite hemispheres of the detector is imposed with $\eta(j_{1})\cdot \eta(j_{2}) < 0$ and $|\Delta \eta_{jj}| > 4.0$. Finally, the invariant mass of the dijet pair, $m_{jj}$, must be greater than 1 TeV.

Similar event selection criteria is used for the $\mu\mu j_{f}j_{f}$ 

\noindent channel. Nevertheless, this channel has significantly lower background contamination in comparison to the  $\tau_{h}\tau_{h} j_{f}j_{f}$ channel. This is due to a much smaller jet$\to\mu$ misidentification rate (at least one order of magnitude smaller). 
The low jet$\to\mu$ misidentification rate allows for lower $p_{T}$ thresholds on $\mu$ and jet candidates. Muons and jets are required to have $p_{T} > 30$ GeV. The less stringent VBF selection in the $\mu\mu j_{f}j_{f}$ channel allows us to regain signal acceptance. Although we have assumed the ATLAS and CMS dimuon triggers allow for 30 GeV cuts on $p_{T}(\mu)$ (mostly to allow for observation of a $Z$ mass peak), it should be noted that the average transverse momentum of muons from $Z'$ decays is $\sim m(Z')$/2, independent of the $Z'$ boost from the VBF selection, and thus the muon $p_{T}$ thresholds can be increased without affecting the projected signal significance. We also note the efficiencies of the central and VBF selections outlined in Table~\ref{tab:selections} are compared with similar results from CMS and agree within 10\%.

Other sets of topological variables were considered, such as the cos$\Delta\phi(\tau_{h},\tau_{h})$ and $\zeta$ variables used in the ATLAS and CMS searches to reduce the W+jets background and targeting the back-to-back nature of the lepton pair~\cite{ATLASZprimeTauTau,CMSZprimeTauTauPAS}. However, since the $Z'$ produced in VBF processes is boosted in order to balance the momentum of the VBF jets, those variables are sub-optimal in these 
studies. On the other hand, the W+jets background is greatly reduced owing to the VBF criteria.

The reconstructed dilepton mass is proposed as the main variable to discriminate against known SM backgrounds.  An enhancement in the tails of the mass distribution would indicate the presence of new physics at

 \begin{figure}[H]
 \begin{center} 
 \includegraphics[width=0.5\textwidth, height=0.35\textheight]{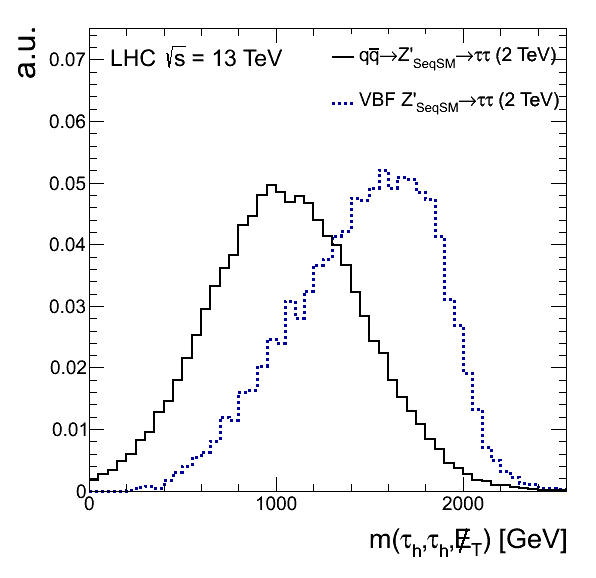}
 \end{center}
 \caption{A comparison of the $m(\tau_{h},\tau_{h},p_{T}^{miss})$ distributions in DY and VBF production of $Z'\to\tau\tau$ with $m(Z')=2$ TeV.}
 \label{fig:DYandVBFditaumass}
 \end{figure}

\noindent the LHC. While the reconstructed dimuon mass $m(\mu,\mu)$ in VBF $Z'\to\mu\mu$ shows up as a narrow bump, the visible ditau mass (i.e. $m(\tau_{h},\tau_{h})$) in VBF $Z'\to\tau\tau$ does not produce a narrow "peak" since the $\tau\tau$ system decays into neutrinos. For this reason, the reconstructed invariant mass distribution in the $\tau_{h}\tau_{h} j_{f}j_{f}$ channel, $m(\tau_{h},\tau_{h},p_{T}^{miss})$, utilizes the reconstructed $p^{miss}_{T}$ and is defined based on Equation~\ref{eq:mT}.

\begin{equation}
\sqrt{(E_{\tau_{h,1}}+E_{\tau_{h,2}}+p_{T}^{miss})^{2}+( \vec{p}_{\tau_{h,1}}+\vec{p}_{\tau_{h,2}}+\vec{p}_{T}^{miss})^{2}}
\label{eq:mT}
\end{equation}

\begin{figure}[H]
 \begin{center} 
 \includegraphics[width=0.5\textwidth, height=0.35\textheight]{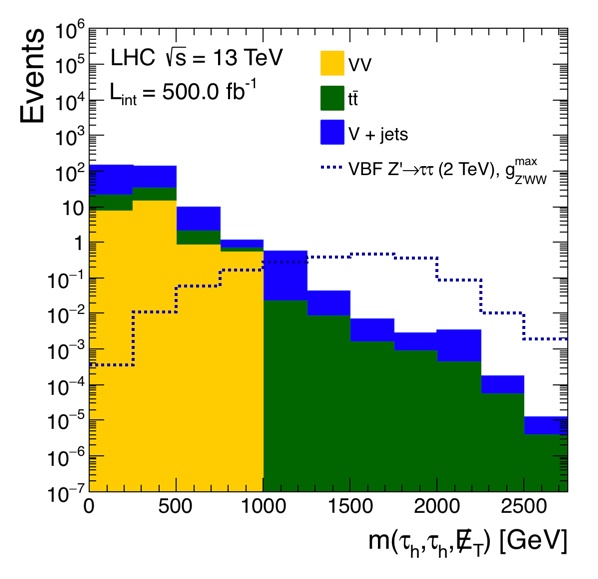}
 \end{center}
 \caption{$m(\tau_{h},\tau_{h},p_{T}^{miss})$ distribution for the main backgrounds and chosen signal benchmark point, after applying the final event selection criteria.}
 \label{fig:massDiTau}
 \end{figure}

\begin{figure}[H]
 \begin{center} 
 \includegraphics[width=0.5\textwidth, height=0.35\textheight]{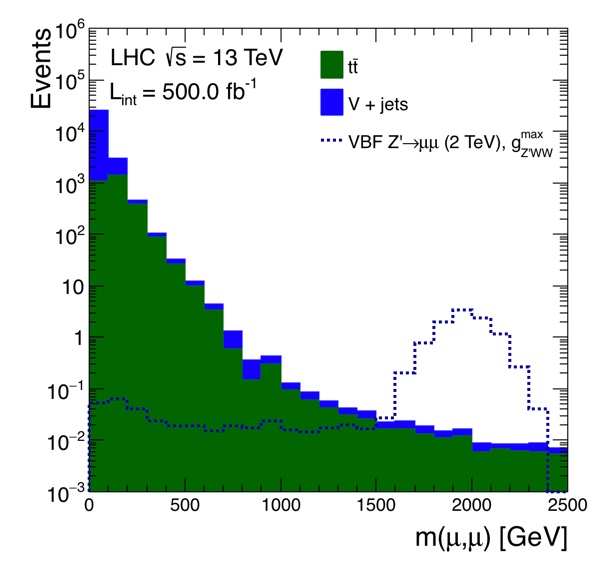}
 \end{center}
 \caption{$m(\tau_{h},\tau_{h},p_{T}^{miss})$ distribution for the main backgrounds and chosen signal benchmark point, after applying the final event selection criteria.}
 \label{fig:massDiMuon}
 \end{figure}

Figure~\ref{fig:DYandVBFditaumass} shows the $m(\tau_{h},\tau_{h},p_{T}^{miss})$ distributions in DY and VBF production of $Z'\to\tau\tau$ with $m(Z')=2$ TeV. 
Since the neutrinos from the $\tau\tau\to\tau_{h}\tau_{h}\nu_{\tau}\nu_{\tau}$ decay are back-to-back in DY processes, there is a cancellation of missing momentum which does not allow the reconstruction of the true mass, on average (i.e. $\langle m(\tau_{h},\tau_{h},p_{T}^{miss}) \rangle \sim 1$ TeV in $q\bar{q}\to Z' \to \tau\tau$). On the other hand, and as noted above, since the $Z'$ produced in VBF processes is boosted in order to balance the momentum of the VBF jets, the neutrinos are not back-to-back and thus results in improved reconstructed mass scale and resolution. 
Figure \ref{fig:massDiTau} shows the expected $m(\tau_{h},\tau_{h},p_{T}^{miss})$ background and signal distributions using events satisfying the $\tau_{h}\tau_{h} j_{f}j_{f}$ cuts in Table~\ref{tab:selections}. The signal is overlaid on top of the stacked 
backgrounds. Figure~\ref{fig:massDiMuon} shows a similar distribution for the $\mu\mu$ channel. The last bin in these distributions represents the overflow bin. The bulk of the background distribution resides at low mass, while the signal dominates in the tails of the distribution.  Although the $VV$ background in Figure 6 lacks statistics above 1 TeV, we have cross-checked that it is

\noindent indeed negligible (upper limit on the yield is 10$^{-3}$ beyond 1 TeV).

\section{Results}

Similar to the ATLAS and CMS searches, we utilize a shape based analysis of the reconstruced mass distribution, using the ROOFit \cite{ROOTFit} toolkit, to construct a binned likelihood following the test statistic based on the profile likelihood ratio. Reasonable systematic uncertainty is considered in order to calculate projected significance. However, since the background rates are small at high mass values where the signal dominates, the poisson uncertainty on the yields dominate the total uncertainty. The dominant sources of systematics are expected to be the uncertainty on $\tau_{h}$ identification (6\% based on \cite{CMSTauID}), 
VBF selection efficiency (20\% based on~\cite{VBF2,CMSVBFDM}), and the uncertainty due to the variations in the yields and mass shapes arising from the choice of parton distribution function (15\% based on~\cite{ATLASZprimeMuMuPAS,CMSZprimeMuMuPAS,ATLASZprimeTauTau,CMSZprimeTauTauPAS}). Based on these considerations, a 25\% total systematic uncertainty on the signal and background yields is a reasonable choice. Systematic uncertainties are incorporated via nuisance parameters following the frequentist approach. A local p-value is calculated as the probability under a background only hypothesis to obtain a value of the test statistic as large as that obtained with a signal plus background hypothesis. The significance $S$ is then determined as the value at which the integral of a Gaussian between $S$ and $\infty$ results in a value equal to the local p-value. 

\begin{figure}[H]
 \begin{center} 
 \includegraphics[width=0.5\textwidth, height=0.35\textheight]{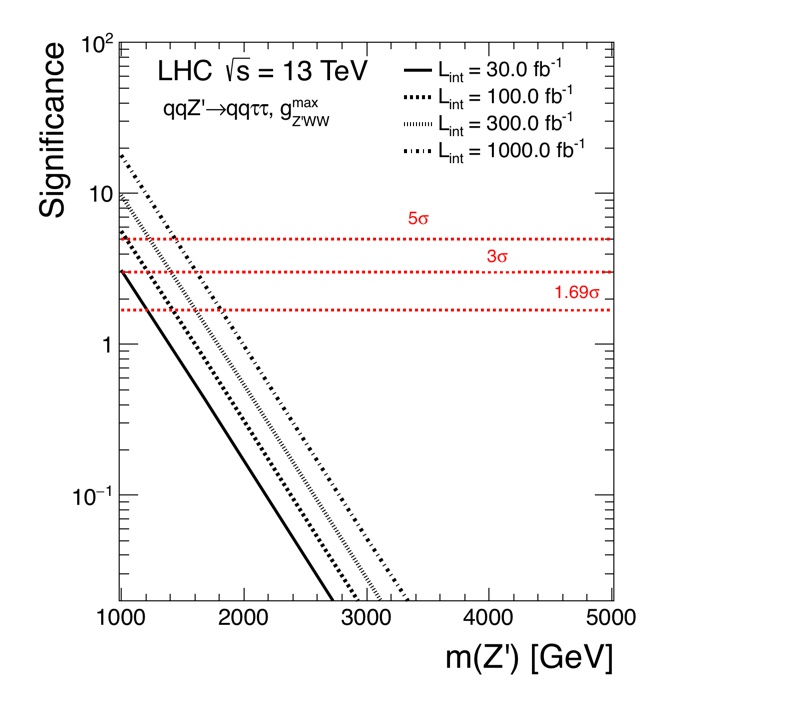}
 \end{center}
 \caption{Projected signal significance for the proposed VBF $Z'\to\tau\tau$ search, using a shape based statistical analysis of the reconstructed mass distribution, as a function of $m(Z')$. The assumed coupling is $g_{Z'VV} = g_{Z'VV}^{max}$. A total systematic uncertainty of 25\% has been considered on the signal and background yields.}
 \label{fig:results}
 \end{figure}

Figure~\ref{fig:results} shows the expected signal significance, considering integrated luminosities between 30 and 1000 fb$^{-1}$, and assuming $g_{Z'VV} = g_{Z'VV}^{max}$. The proposed $Z'$ searches using the VBF topology can provide expected significances greater than 5$\sigma$ (3$\sigma$) for $Z'$ masses up to 1.4 (1.6) TeV and 2.0 (2.2) TeV in the $\tau\tau j_{f}j_{f}$ and $\mu\mu j_{f}j_{f}$ channels, respectively. Using 1.69$\sigma$ to define an exclusion bound at 95\% confidence level, the projected exclusion bounds are $m(Z') < 1.8$ TeV and $m(Z') < 2.5$ TeV in the $\tau_{h}\tau_{h} j_{f}j_{f}$ and $\mu\mu j_{f}j_{f}$ channels. Figures 10-12 show the ratio of VBF $Z'$ to DY $Z'$ signal significances, $S_{VBF} / S_{DY}$, as a function of $\kappa_{q}$ and $\kappa_{V}$. For fixed $\kappa_{V} = 0.25$, $0.5$, and 1, the modifier $\kappa_{q}$ varies between 0.1 and 1. The VBF $Z'$ search is the most important mode for discovery when $\kappa_{q} \le 0.3$ (0.2) for $g_{Z'VV} = g_{Z'VV}^{max}$ ($g_{Z'VV} = \frac{1}{2} g_{Z'VV}^{max}$). There exist models with $\kappa_{q} \le 0.3$ (e.g.~\cite{ZpModel2, ZpModel1}), including those constructed to explain the B-meson anomalies observed at LHCb~\cite{LHCb}. 
We stress that while it's clear the $\mu\mu j_{f}j_{f}$ channel has better discovery potential for models with universal couplings of the $Z'$ boson to the leptons, the $\tau\tau j_{f}j_{f}$ channel is particularly important in models with enhanced couplings to third-generation fermions (e.g. topcolor-assisted technicolor~\cite{TAT}) and can be derived from Figure~\ref{fig:results} by appropriately rescaling the VBF $Z'$ cross-section.

\section{Discussion}

The main result of this paper is that probing heavy neutral gauge bosons $Z'$ produced through VBF processes can be a key methodology to complement current

 \begin{figure}[H]
 \begin{center} 
 \includegraphics[width=0.5\textwidth, height=0.35\textheight]{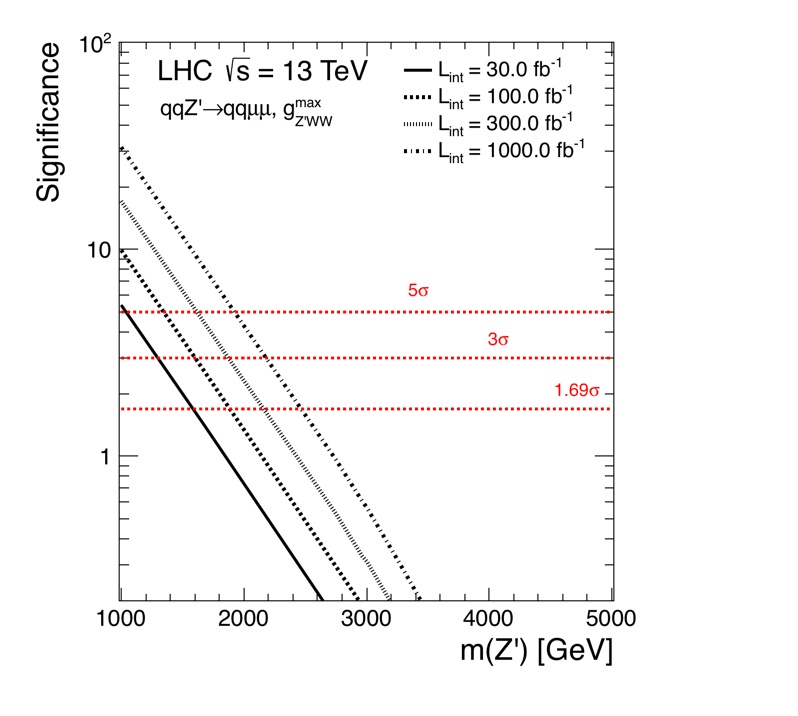}
 \end{center}
 \caption{Projected signal significance for the proposed VBF $Z'\to\mu\mu$ search, using a shape based statistical analysis of the reconstructed mass distribution, as a function of $m(Z'_{SeqSM})$. The assumed coupling is $g_{Z'VV} = g_{Z'VV}^{max}$. A total systematic uncertainty of 25\% has been considered on the signal and background yields.}
 \label{fig:resultsb}
 \end{figure}
 
 \begin{figure}[H]
 \begin{center} 
 \includegraphics[width=0.5\textwidth, height=0.35\textheight]{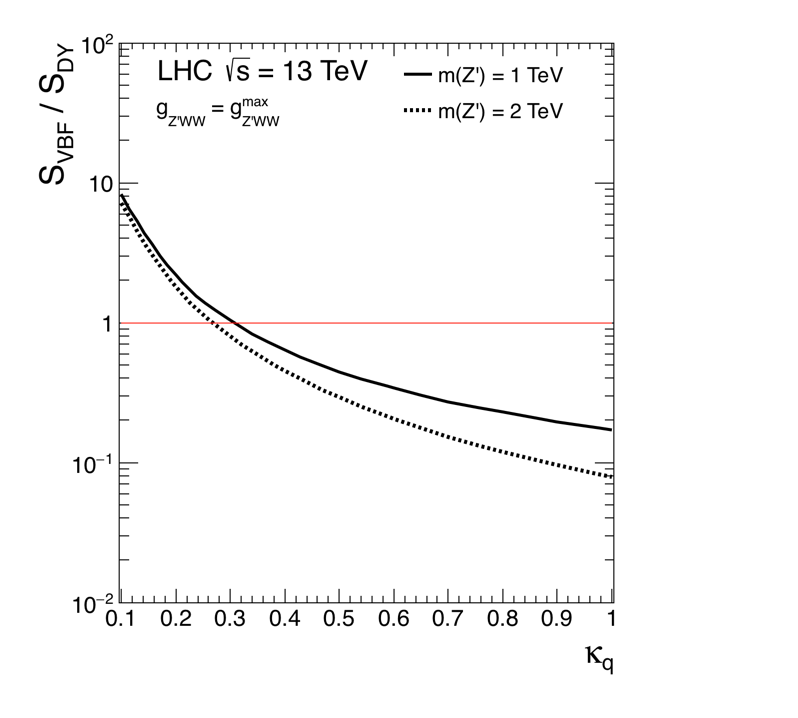}
 \end{center}
 \caption{The ratio of VBF $Z'$ to DY $Z'$ signal significances, as a function of $\kappa_{q}$ and for fixed $\kappa_{V} = 1$.}
 \label{fig:resultsc}
 \end{figure}
 
\noindent searches for TeV scale particles at the LHC. In the context of a simplified phenomenological approach, it is shown that the VBF search sensitivity surpasses that of the more standard DY $Z'$ search at approximately $\kappa_{q} \le 0.3$, which makes $Z'$ searches with the VBF topology important for the long term search strategies at the LHC. This is also particularly interesting since there has been no evidence for a TeV scale $Z'$ at the LHC. The $Z'$ boson may have eluded the CMS and ATLAS experiments because $g_{Z'qq}$ is small, and thus would not be discoverable by the ``standard" DY $Z'$ searches conducted thus far. Additionally, even if a $Z'$ boson is discovered in the DY search channel when $\kappa_{q} \approx 1$, a VBF $Z'$ search remains a key part of the $Z'$ search program at the LHC in order to establish the couplings of the $Z'$ to the SM vector bosons. To highlight the power of the VBF topology in $Z'$ searches with large QCD backgrounds as well as those with clean signatures, we consider $Z'$ decays to $\tau\tau$ and $\mu\mu$ and show that  the requirement of a dilepton pair combined with 
two high $p_{T}$ forward jets with large separation in pseudorapidity and with large dijet mass is effective in reducing SM backgrounds. The expected exclusion bounds (at 95\% confidence level) are $m(Z') < 1.8$ TeV and $m(Z') < 2.5$ TeV in the $\tau_{h}\tau_{h} j_{f}j_{f}$ and $\mu\mu j_{f}j_{f}$ channels, respectively, assuming 1000 fb$^{-1}$ of 13 TeV data from the LHC and $g_{Z'VV} = g_{Z'VV}^{max}$. The use of the VBF topology to search for massive neutral gauge bosons at the LHC produces expected significances greater than 5$\sigma$  (3$\sigma$) for $Z'$ masses up to 1.4 (1.6) TeV and 2.0 (2.2) TeV in the $\tau_{h}\tau_{h} j_{f}j_{f}$ and $\mu\mu j_{f}j_{f}$ channels. 

 \begin{figure}[H]
 \begin{center} 
 \includegraphics[width=0.5\textwidth, height=0.35\textheight]{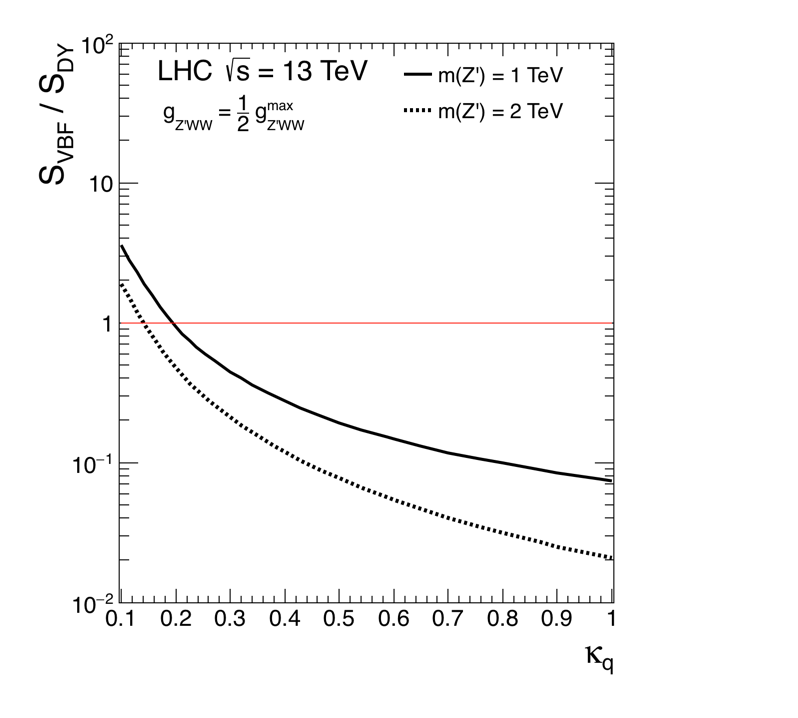}
 \end{center}
 \caption{The ratio of VBF $Z'$ to DY $Z'$ signal significances, as a function of $\kappa_{q}$ and for fixed $\kappa_{V} = 0.5$.}
 \label{fig:resultsd}
 \end{figure}

 \begin{figure}[H]
 \begin{center} 
 \includegraphics[width=0.5\textwidth, height=0.35\textheight]{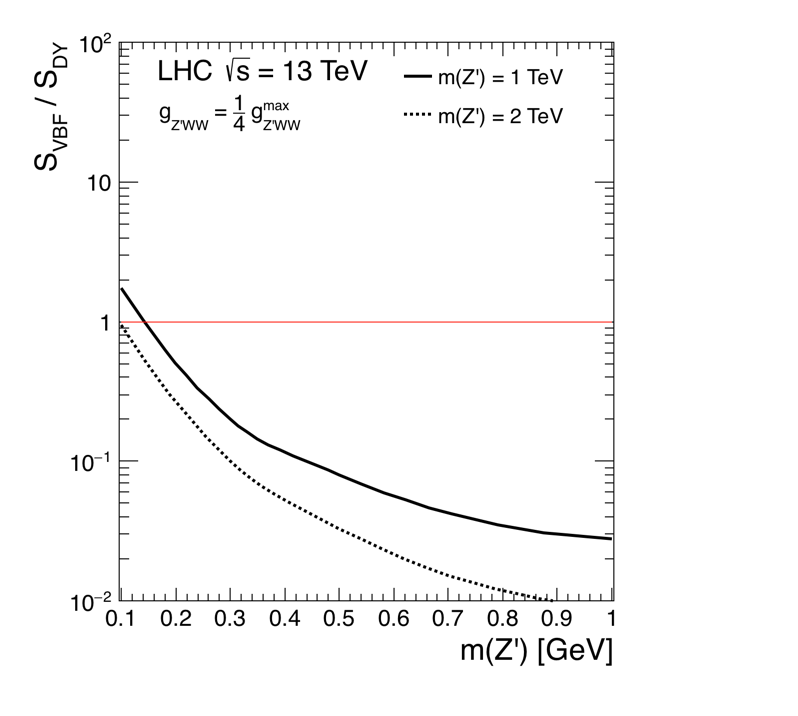}
 \end{center}
 \caption{The ratio of VBF $Z'$ to DY $Z'$ signal significances, as a function of $\kappa_{q}$ and for fixed $\kappa_{V} = 0.25$.}
 \label{fig:resultsd}
 \end{figure}
  
\section{Acknowledgements}

We thank the constant and enduring financial support received for this project from the faculty of science at Universidad de los Andes (Bogot\'a, Colombia), the administrative department of science, technology and innovation of Colombia (COLCIENCIAS), the Physics \& Astronomy department at Vanderbilt University and the US National Science Foundation. This work is supported in part by NSF Award PHY-1506406. TW thanks DoE for partial support via grant DE-SC0011981.

\end{document}